\documentclass[a4paper,12pt]{article}
\usepackage{amssymb}
\newtheorem{thm}{Theorem}[section]
\newtheorem{prop}[thm]{Proposition}
\newtheorem{lemma}[thm]{Lemma}

\def\o#1{\overline{#1}}

\def\P{{\mathbb P}}

\def\qed{$\square$}
\def\dfrac#1#2{{\displaystyle\frac{#1}{#2}}}
\begin{document}

\begin{center}
{\Large {\bf Lax formalism for $q$-Painlev\'e equations
with affine Weyl group symmetry of type $E^{(1)}_n$}}\\
\vskip10mm
{\large Yasuhiko Yamada}\\
\vskip5mm
Department of Mathematics, Faculty of Science,\\
Kobe University, Hyogo 657-8501, Japan
\end{center}

\vskip10mm

\noindent
{\bf Abstract.}
An explicit form of the Lax pair for the $q$-difference Painlev\'e equation with 
affine Weyl group symmetry of type $E^{(1)}_8$ is obtained.
Its degeneration to $E^{(1)}_7$, $E^{(1)}_6$ and  $D^{(1)}_5$ cases are also given.

\vskip5mm



\vskip10mm

\section{Introduction}
A Lax formalism for the elliptic difference Painlev\'e equation \cite{Sakai} was obtained in \cite{e8ell}.
The construction is concrete including the specification of the unknown variable $(f,g)$ of the Painlev\'e equation. 
The explicit formula of the elliptic Painlev\'e equation and its Lax from, however, 
seems to be too complicated to write it down in tidy form.

On the other hand, for the multiplicative (i.e. $q$-difference) and additive cases, rather concise expressions
for the Painlev\'e equations have been known \cite{ORG}.
The aim of this paper is to adjust the construction of \cite{e8ell} to the $q$-Painlev\'e equations in \cite{ORG} and
write down the corresponding Lax equations explicitly. 
We note that the Lax formulations of difference Painlev\'e equations were obtained in \cite{AB} for additive $E^{(1)}_6$,
\cite{B} for additive $E^{(1)}_7$, $E^{(1)}_8$ and \cite{SakaiA2} for $q$-$E^{(1)}_6$ cases (see also \cite{Sakai-rims}).

In general, the scalar Lax pair consists of a linear (difference) equation $L_1=0$, and its deformation
equation $L_2=0$. Though the Lax equations $L_1, L_2$ in \cite{e8ell} were both of degree $(3,2)$ in variables $(f,g)$, 
other choices of equation $L_2$ are possible depending on the direction of deformation.
In this paper, we will give an explicit expressions of the Lax pair $L_1, L_2$ 
for $q$-Painlev\'e equation (\ref{eq:T-bar}) corresponding to a distinguished direction. 
The first equation $L_1$ (\ref{eq:L1}) is of degree $(3,2)$ which is the explicit realization of that
in \cite{e8ell}. The second equation $L_2$ (\ref{eq:L2}) is of degree $(1,1)$ and hence much more economical.

The contents of this paper is as follows.
In section \ref{sect:basic}, the explicit expression of the $q$-Painlev\'e equation of 
type $E^{(1)}_8$ is recapitulated. (Its Weyl group symmetry is explained in Appendix A.)
In section \ref{sect:lax}, corresponding Lax equations are given in explicit form.
The compatibility is proved in section \ref{sect:proof}.
Finally, in section \ref{sect:dege}, the degeneration to $E^{(1)}_7$, $E^{(1)}_6$ and
$D^{(1)}_5$ cases are considered. The Lax form for $q$-$E^{(1)}_7$ case seems also to be new.

\section{Fundamental equations}\label{sect:basic}
Let $h_1, h_2, u_1, \ldots, u_8$ and $q=h_1^2h_2^2/(u_1\cdots u_8)$ be complex parameters.
We consider a configuration of the following eight points in $\P^1 \times \P^1$:
\begin{equation}
P_i=P(u_i), \quad P(u)=(f(u),g(u)), \quad f(u)=u+\dfrac{h_1}{u}, \ g(u)=u+\dfrac{h_2}{u}.
\end{equation}
The functions $f=f(u), g=g(u)$ give a parametrization of a rational curve $C_0 : \varphi(f, g)=0$ of degree $(2,2)$, where
\begin{equation}
\varphi(f,g)=(f-g)(\dfrac{f}{h_1}-\dfrac{g}{h_2})-(h_1-h_2)(\dfrac{1}{h_1}-\dfrac{1}{h_2}).
\end{equation}

We define a polynomial $U(z)$ as
\begin{equation}
U(z)=\prod_{i=1}^8(z-u_i)=\sum_{i=0}^8 (-1)^im_{8-i}z^i.
\end{equation}
Hence $m_0=1$, $m_8=h_1^2h_2^2/q$. We also define $P_n(h,x), P_d(h,x)$ by
\begin{equation}\label{eq:Prel1}
\begin{array}l
(z-\dfrac{h}{z})P_n(h,z+\dfrac{h}{z})=\dfrac{1}{z^3}U(z)-(\dfrac{z}{h})^3U(\dfrac{h}{z}),\\[4mm]
(z-\dfrac{h}{z})P_d(h,z+\dfrac{h}{z})={z^5}U(\dfrac{h}{z})-(\dfrac{h}{z})^5U(z).
\end{array}
\end{equation}
Then they satisfies
\begin{equation}\label{eq:Prel2}
\begin{array}l
P_d(h,z+\dfrac{h}{z})+h^3 z^2 P_n(h,z+\dfrac{h}{z})-(\dfrac{h}{z})^3(z+\dfrac{h}{z})U(z)=0, \\[4mm]
P_d(h,g)=h^4P_n(\dfrac{1}{h},\dfrac{g}{h})|_{m_i \mapsto m_{8-i}}.
\end{array}
\end{equation}
Explicitly, we have
\begin{equation}\label{eq:Prel3}
\begin{array}l
P_n(h,g)= m_0g^4-m_1g^3+(m_2-3hm_0-h^{-3}m_8)g^2\\
\qquad +(2hm_1-m_3+h^{-2}m_7)g+(h^2m_0-hm_2+m_4-h^{-1}m_6+h^{-2}m_8),\\[4mm]
P_d(h,g)= m_8g^4-hm_7g^3+(h^2m_6-3hm_8-h^{5}m_0)g^2\\
\qquad +(2h^2m_7-h^3m_5+h^{5}m_1)g+(h^6m_0-h^5m_2+h^4m_4-h^{3}m_6+h^{2}m_8).
\end{array}
\end{equation}

The $q$-Painlev\'e equation of type $E^{(1)}_8$ can be described by the
following bi-rational transformations \cite{ORG}:
\begin{equation}\label{eq:T-bar}
T: (h_1,h_2, u_1,\ldots, u_8;f,g)\mapsto (\dfrac{h_1}{q}, h_2 q, u_1,\ldots, u_8;\o{f},\o{g}), 
\end{equation}
where
\begin{equation}\label{eq:fueq}
\dfrac{(\o{f}-g)(f-g)-(\frac{h_1}{q}-h_2)(h_1-h_2)\frac{1}{h_2}}
{(\frac{\o{f}q}{h_1}-\frac{g}{h_2})(\frac{f}{h_1}-\frac{g}{h_2})
-(\frac{q}{h_1}-\frac{1}{h_2})(\frac{1}{h_1}-\frac{1}{h_2})h_2}
=\dfrac{h_1^2h_2^4}{q}\dfrac{P_n(h_2,g)}{P_d(h_2,g)},
\end{equation}
and 
\begin{equation}\label{eq:gueq}
\dfrac{(\o{f}-\o{g})(\o{f}-g)-(\frac{h_1}{q}-h_2 q)(\frac{h_1}{q}-h_2)\frac{q}{h_1}}
{(\frac{\o{f}q}{h_1}-\frac{\o{g}}{h_2 q})
(\frac{\o{f}q}{h_1}-\frac{g}{h_2})-(\frac{q}{h_1}
-\frac{1}{h_2 q})(\frac{q}{h_1}-\frac{1}{h_2})\frac{h_1}{q}}
=\dfrac{h_1^4h_2^2}{q^3}\dfrac{P_n(\frac{h_1}{q},\o{f})}{P_d(\frac{h_1}{q},\o{f})}.
\end{equation}

Under the transformation
\begin{equation}
f\leftrightarrow \o{g}, \quad
g\leftrightarrow \o{f}, \quad
h_1 \rightarrow h_2 q, \quad
h_2 \rightarrow \frac{h_1}{q},
\end{equation}
the equations (\ref{eq:fueq}) and (\ref{eq:gueq}) are transformed with each other.

Define polynomial $V=V(f_0, f)$ as
\begin{equation}\label{eq:Vpoldef}
\begin{array}l
V(f_0,f)=q\Big[(f_0-g)(f-g)-(\dfrac{h_1}{q}-h_2)(h_1-h_2)\dfrac{1}{h_2}\Big]P_d(h_2,g)\\
-h_1^2h_2^4\Big[(\dfrac{f_0 q}{h_1}-\dfrac{g}{h_2})(\dfrac{f}{h_1}-\dfrac{g}{h_2})
-(\dfrac{q}{h_1}-\dfrac{1}{h_2})(\dfrac{1}{h_1}-\dfrac{1}{h_2}){h_2}\Big]P_n(h_2,g).
\end{array}
\end{equation}
Then eq.(\ref{eq:fueq}) is written as 
\begin{equation}\label{eq:of-f}
V(\o{f},f)=0.
\end{equation}

In the following, we also use the notation
\begin{equation}
\o{f}(u)=u+\dfrac{h_1}{qu}, \quad
\o{g}(u)=u+\dfrac{h_2 q}{u}.
\end{equation}

\section{Lax equations}\label{sect:lax}

Let us define the Lax pair for the $q$-Painlev\'e equations (\ref{eq:fueq}), (\ref{eq:gueq}).
The first equation $L_1=0$ is a three term $q$-difference equation for $Y(\frac{z}{q}), Y(z), Y(qz)$
defined by
\begin{equation}\label{eq:L1}
\begin{array}l
L_1=
\dfrac{q^5U(\frac{z}{q})}{(z^2-h_1 q^2)\{f-f(\frac{z}{q})\}}\Big[Y(\frac{z}{q})-\dfrac{g-g(\frac{h_1 q}{z})}{g-g(\frac{z}{q})}Y(z)\Big]\\[6mm]
\qquad +\dfrac{z^8U(\frac{h_1}{z})}{(z^2-h_1)h_1^4\{f-f(z)\}}\Big[Y(qz)-\dfrac{g-g(z)}{g-g(\frac{h_1}{z})}Y(z)\Big]\\[6mm]
\qquad +\dfrac{(h_1-h_2)z^2(z^2-h_1 q)V(\o{f}(\frac{z}{q}),f)}{h_1^3 h_2^3 q g \varphi  \{g-g(\frac{h_1}{z})\}\{g-g(\frac{z}{q})\}}Y(z),
\end{array}
\end{equation}
where $V$ is in eq.(\ref{eq:Vpoldef}).
The following proposition shows that the equation $L_1=0$ has the geometric properties described
in \cite{e8ell}.
\begin{prop}\label{prop1}
Put $F(f,g)=\varphi\{f-f(\frac{z}{q})\}\{f-f(z)\} L_1$. 
Then the algebraic curve $F=0$ in $\P^1\times \P^1$ satisfy the following properties:\\
(i) It is of degree $(3,2)$.\\
(ii) It passes the 12 points $P_1, \ldots, P_8$, $P(z)$, $P(\frac{h_1 q}{z})$, $Q(z)$ and $Q(\frac{z}{q})$, where
$Q(u)=(f,g)$ is defined by $f=f(u)$ and $\dfrac{g-g(u)}{g-g(\frac{h_1}{u})}=\dfrac{Y(qu)}{Y(u)}$,  
for $u=z, \frac{z}{q}$.\\
Moreover, these conditions determine the curve $F=0$ uniquely.
\end{prop}

\noindent
{\it Proof.}
(i) It is easy to see that the coefficients
of $Y(qz)$ and $Y(\frac{z}{q})$ in $F$ are polynomials in $(f, g)$ of degree $(3,2)$.
The coefficient of $Y(z)$ looks a rational function with numerator of degree $(3,6)$ and denominator of degree $(0,3)$.
However its residues at $g=0$, $g=g(\frac{z}{q})$, $g=g(\frac{h_1}{z})$
and leading term at $g \rightarrow \infty$ are proportional to
\begin{equation}
\begin{array}{lll}
P_d(h_2,0)-h_2^4 P_n(h_2,0), 
&{\rm at}&\ g=0\\[2mm]
P_d(h_2,g(\frac{z}{q}))+h_2^3 (\frac{z}{q})^2 P_n(h_2,g(\frac{z}{q}))-(\frac{h_2 q}{z})^3g(\frac{z}{q})U(\frac{z}{q}), 
&{\rm at}&\ g=g(\frac{z}{q})\\[2mm]
P_d(h_2,g(\frac{h_1}{z}))+h_2^3 (\frac{h_1}{z})^2 P_n(h_2,g(\frac{h_1}{z}))-(\frac{h_2 z}{h_1})^3g(\frac{h_1}{z})U(\frac{h_1}{z}), 
&{\rm at}&\ g=g(\frac{h_1}{z})\\[2mm]
(h_1^2h_2^2 m_0-q m_8)g^3, 
&{\rm at}&\ g\rightarrow \infty
\end{array}
\end{equation}
and all of them are zero due to eqs (\ref{eq:Prel2}), (\ref{eq:Prel3}). Hence the property (i) is proved.

(ii) The coefficients of $Y(qz)$ and $Y(\frac{z}{q})$ of polynomial $F$ trivially vanish at $P_1, \ldots, P_8$, $P(z)$ and $P(\frac{h_1 q}{z})$.
The coefficient of $Y(z)$ is proportional to
\begin{equation}\label{eq:Ycoef}
(u-z)(h_1 q-u z)\{P_d(h_2,g(u))+h_2^3 u^2 P_n(h_2,g(u))\}=(u-z)(h_1 q-u z)(\frac{h_2}{u})^3 g(u)U(u)
\end{equation}
under the specialization $f=f(u)$ and $g=g(u)$. Hence, it also vanishes at $P_1, \ldots, P_8$, $P(z)$ and $P(\frac{h_1 q}{z})$.
The vanishing of $F$ at $Q(z)$ and $Q(\frac{z}{q})$ can be directly seen from the structure of $L_1$.
The property (ii) is proved.

The uniqueness follows from simple dimensional argument.\qed

In \cite{e8ell}, the second Lax equation $L_2=0$ is also defined as a curve of degree $(3,2)$ with
similar vanishing conditions. However, the $L_2$ equation (the deformation equation) depends on the 
direction in $E^{(1)}_8$-translations and the one in \cite{e8ell} is not what we want here. In fact, we can take
more simple Lax equation $L_2=0$ given by
\begin{equation}\label{eq:L2}
\begin{array}l
L_2=\{g-g(\frac{z}{q})\}Y(\frac{z}{q})-\{g-g(\frac{h_1 q}{z})\}Y(z)+\{f-f(z)\}(\dfrac{h_1}{z}-\dfrac{z}{q^2})\o{Y}(\frac{z}{q}).
\end{array}
\end{equation}

The following is the main result of this paper.
\begin{thm}\label{thm:main}
The compatibility of the equations $L_1=0$ (\ref{eq:L1}) and $L_2=0$ (\ref{eq:L2})
gives the Painlev\'e equation (\ref{eq:fueq}), (\ref{eq:gueq}).
\end{thm}

\section{Proof of the main theorem}\label{sect:proof}

Here, we will prove the Theorem.\ref{thm:main}.

Using $L_2=0$ (\ref{eq:L2}) and its shift $L_2|_{z \rightarrow q z}=0$, eliminate $Y(qz)$ and $Y(\frac{z}{q})$ from
$L_1=0$ (\ref{eq:L1}), we get
\begin{equation}\label{eq:elim1}
\begin{array}l
L_1=
\dfrac{q^3 U(\frac{z}{q})}{z\{g-g(\frac{z}{q})\}}\o{Y}(\frac{z}{q})
-\dfrac{z^7 U(\frac{h_1}{z})}{h_1^4 q \{g-g(\frac{h_1}{z})\}}\o{Y}(z)\\[6mm]
\qquad +\dfrac{(h_1-h_2)z^2(z^2-h_1 q)V(\o{f}(\frac{z}{q}),f)}{h_1^3 h_2^3 q g \varphi  \{g-g(\frac{h_1}{z})\}\{g-g(\frac{z}{q})\}}Y(z)=0.
\end{array}
\end{equation}
We introduce auxiliary variable $W(z)$ by
\begin{equation}\label{eq:w2y}
W(\frac{z}{q})=\o{Y}(\frac{z}{q})-\dfrac{z^8}{h_1^4 q^4}\dfrac{\{g-g(\frac{z}{q})\}}{\{g-g(\frac{h_1}{z})\}}\dfrac{U(\frac{h_1}{z})}{U(\frac{z}{q})}\o{Y}(z),
\end{equation}
then, from eq. (\ref{eq:elim1}) we have
\begin{equation}\label{eq:wy0}
\dfrac{q^3 U(\frac{z}{q})}{z\{g-g(\frac{z}{q})\}}W(\frac{z}{q})
+\dfrac{(h_1-h_2)z^2(z^2-h_1 q)V(\o{f}(\frac{z}{q}),f)}{h_1^3 h_2^3 q g \varphi  \{g-g(\frac{h_1}{z})\}\{g-g(\frac{z}{q})\}}Y(z)=0.
\end{equation}
From eq.(\ref{eq:wy0}), (\ref{eq:wy0})$|_{z \rightarrow q z}$ and $L_2|_{z \rightarrow q z}$, we eliminate $Y(z)$ and $Y(qz)$ and obtain
\begin{equation}\label{eq:wwy1}
\begin{array}l
W(\frac{z}{q})
+\dfrac{({h_1}-{h_2}) ({h_1}-z^2) ({h_1} q-z^2) (f-f(z)) V(\o{f}(\frac{z}{q}),f) z^2}
{g {h_1}^3 {h_2}^3 {\varphi} q^5 (g-g(\frac{{h_1}}{z})) (g-g(z))U(\frac{z}{q})}\o{Y}(z)\\[6mm]
+\dfrac{({h_1} q-z^2) (g-g(\frac{{h_1}}{q z})) U(z)V(\o{f}(\frac{z}{q}),f)}{q^4 (q z^2-{h_1}) (g-g(z)) U(\frac{z}{q}) V(\o{f}(z),f)}
W(z)=0.
\end{array}   
\end{equation}

\begin{lemma}
We have
\begin{equation}\label{eq:x1x2}
\dfrac{(f-x_1)V(x_2,f)}{(\o{f}-x_2)V(\o{f},x_1)}=\dfrac{(h_1-h_2 q)\varphi}{(h_1-h_2)\varphi_u},
\end{equation}
where $\varphi_u=\varphi|_{\{f\rightarrow \o{f}, h_1\rightarrow \frac{h_1}{q}\}}$.
\end{lemma}

\noindent
{\it Proof.} The (LHS) of eq.(\ref{eq:x1x2}) is fractional liner function both in $x_1$ and $x_2$.
Whose zeros and poles cancel due to  eq.(\ref{eq:of-f}) hence it is constant.
Taking a limit $x_1, x_2 \rightarrow \infty$, the constant is given by
\begin{equation}
\dfrac{(f-g)P_d(h_2, g) -h_2^3(fh_2-gh_1) P_n(h_2, g)}
{(\o{f}-g)P_d(h_2, g) - h_2^3 (\o{f}h_2-gh_1/q) P_n(h_2, g)}.
\end{equation}
From eq.(\ref{eq:fueq}), this coincide with the (RHS) of eq.(\ref{eq:x1x2}).\qed

By using eq.(\ref{eq:x1x2}), eq.(\ref{eq:wwy1}) can be written as
\begin{equation}\label{eq:wwy2}
\begin{array}l
W(\frac{z}{q})
+\dfrac{({h_1}-{h_2} q) ({h_1}-z^2) ({h_1} q-z^2) (\o{f}-\o{f}(\frac{z}{q})){V}(\o{f},{f}(z)) z^2}
{g {h_1}^3 {h_2}^3 {\varphi_u} q^5 (g-{g}(\frac{{h_1}}{z})) (g-{g}(z))U(\frac{z}{q})}\o{Y}(z)\\[6mm]
+\dfrac{({h_1} q-z^2) (\o{f}-\o{f}(\frac{z}{q}))(g-{g}(\frac{{h_1}}{q z})) U(z)}
{q^4 (q z^2-{h_1}) (\o{f}-\o{f}(z)) (g-{g}(z)) U(\frac{z}{q})}W(z)=0.
\end{array}
\end{equation}
Using eq.(\ref{eq:w2y}), express $W$ in terms of $Y$, we finally obtain
\begin{equation}\label{eq:L1up}
\begin{array}l
L_{1u}=\dfrac{U(\frac{z}{q})}{(z^2-h_1 q^2)\{\o{f}-\o{f}(\frac{z}{q})\}}
\Big[\o{Y}(\frac{z}{q})-\dfrac{z^8}{h_1^4q^4}\dfrac{U(\frac{h_1}{z})}{U(\frac{z}{q})}\dfrac{g-g(\frac{z}{q})}{g-g(\frac{h_1}{z})}\o{Y}(z)\Big]\\[6mm]
\qquad +\dfrac{z^8U(\frac{h_1}{qz})}{(qz^2-h_1)h_1^4\{\o{f}-\o{f}(z)\}}
\Big[\o{Y}(qz)-\dfrac{h_1^4}{q^4 z^8}\dfrac{U(z)}{U(\frac{h_1}{qz})}\dfrac{g-g(\frac{h_1}{qz})}{g-g(z)}\o{Y}(z)\Big]\\[6mm]
\qquad +\dfrac{(h_1-h_2 q)z^2(z^2-h_1)V(\o{f},f(z))}{h_1^3 h_2^3 q^5 g \varphi_u  \{g-g(\frac{h_1}{z})\}\{g-g(z)\}}\o{Y}(z)=0.
\end{array}
\end{equation}
This is the three term difference equation for $\o{Y}(qz), \o{Y}(z), \o{Y}(\frac{z}{q})$ which should be
compared with $T$-evolution of the eq.$L_1$ for the compatibility. To do this, one should make further variable change
from $(\o{f},g)$ to $(\o{f}, \o{g})$, where $g$ and $\o{g}$ are related by eq.(\ref{eq:gueq}). 
Though the explicit computation of this variable change is rather complicated,
one can bypass it by using the geometric method as \cite{e8ell}.

\begin{lemma}\label{lem:L1u}
The algebraic curve $\varphi\{\o{f}-\o{f}(\frac{z}{q})\}\{\o{f}-\o{f}(z)\} L_{1u}=0$
in variables $(\o{f},g) \in \P^1\times \P^1$ is uniquely characterized by the following properties:\\
(i) It is of degree $(3,2)$.\\
(ii) It passes the 10 points $(\o{f},g)=(\o{f}(u),g(u))$ with $u=u_1, \ldots, u_8,\frac{z}{q}, \frac{h_1}{qz}$,
and two more points defined by $\o{f}=\o{f}(u)$ and $\dfrac{q^4 u^8}{h_1^4}\dfrac{U(\frac{h_1}{q u})}{U(u)}\dfrac{g-g(u)}{g-g(\frac{h_1}{qu})}=\dfrac{\o{Y}(u)}{\o{Y}(qu)}$,  
for $u=z, \frac{z}{q}$.
\end{lemma}
\noindent
{\it Proof.}
Since the structure of eq.(\ref{eq:L1up}) is almost the same as eq.(\ref{eq:L1}), the lemma can be
shown in the same way as Proposition.\ref{prop1}.\qed

\begin{lemma}\label{lem:gueq-str}
The bi-rational transformation $\o{g}=\o{g}(\o{f},g)$ [its inverse $g=g(\o{f},\o{g})$, resp.] is uniquely characterized by the
following properties. (i) It is given by a ratio of polynomials of degree $(4,1)$ passing through the 8 points
$(\o{f},g)=(\o{f}(u_i),g(u_i))$ [$(\o{f},\o{g})=(\o{f}(u_i),\o{g}(u_i))$, resp.]. (ii) For generic parameter $u$, it transforms as
\begin{equation}
(\o{f},g)=(\o{f}(\frac{h_1}{qu}),g(\frac{h_1}{qu})) \leftrightarrow (\o{f},\o{g})=(\o{f}(u),\o{g}(u)).
\end{equation}
\end{lemma}
\noindent
{\it Proof.}
The transformation $\o{g}=\o{g}(\o{f},g)$ [$g=g(\o{f}, \o{g})$] is determined by eq.(\ref{eq:gueq}), i.e.
\begin{equation}\label{eq:Vforg}
\begin{array}l
q^3 \left[(\o{f}-\o{g})(\o{f}-g)-(\frac{h_1}{q}-h_2 q)(\frac{h_1}{q}-h_2)\frac{q}{h_1}\right]P_d(\frac{h_1}{q},\o{f})\\[5mm]
-h_1^4h_2^2\left[(\frac{\o{f}q}{h_1}-\frac{\o{g}}{h_2 q})(\frac{\o{f}q}{h_1}-\frac{g}{h_2})
-(\frac{q}{h_1}-\frac{1}{h_2 q})(\frac{q}{h_1}-\frac{1}{h_2})\frac{h_1}{q}\right]P_n(\frac{h_1}{q},\o{f})=0.
\end{array}
\end{equation}
This equation is linear in both $\o{g}$ and $g$, and of degree 4 in $\o{f}$ since degree 5 terms are cancelled.
Hence $\o{g}(\o{f},g)$ [$g(\o{f}, \o{g})$] is of degree (4,1). By a similar computation as eq.(\ref{eq:Ycoef}), 
the eq.(\ref{eq:Vforg}) is written as
\begin{equation}
\{g-g(\frac{h_1}{q u})\}\Big\{q^3 P_d(\frac{h_1}{q},\o{f}(u))+h_1^3 u^2 P_n(\frac{h_1}{q},\o{f}(u))\Big\}=
\{g-g(\frac{h_1}{q u})\}(\frac{h_1}{u})^3\o{f}(u)U(u)=0,
\end{equation}
when $\o{f}=\o{f}(u)$ and $\o{g}=\o{g}(u)$. Then it follows that  (1) if $\o{f}=\o{f}(u_i)$ then  $\o{g}=\o{g}(u_i)$ (regardless of $g$)
and (2) if $\o{f}=\o{f}(u)$ and $\o{g}=\o{g}(u)$ then $g=g(\frac{h_1}{q u})$ for $u \neq u_i$. 
Hence the properties (i), (ii) are verified. Finally the uniqueness follows from dimensional argument.\qed

\begin{lemma}\label{lem:gog}
Let $F(\o{f},g)=0$ be any curve of degree $(3,2)$ passing through the 10 points in (ii) of Lemma.\ref{lem:L1u}.
Then the curve in variables $(\o{f}, \o{g})$ obtained from $F(\o{f}, g(\o{f},\o{g}))=0$ is also of degree $(3,2)$ and
passes the 10 points $(\o{f},\o{g})=(\o{f}(u),\o{g}(u))$ with $u=u_1, \ldots, u_8, z, \frac{h_1}{z}$.
\end{lemma}
\noindent
{\it Proof.}
From lemma \ref{lem:gueq-str}, we have $g=\frac{A_{41}}{B_{41}}$ where $A_{41}, B_{41}$ are polynomials of degree $(4,1)$ in
$(\o{f}, \o{g})$ vanishing at the 8 points : $(\o{f}(u_i), \o{g}(u_i))_{i=1}^8$. Substitute this into $F(\o{f}, g)$ we have
$F(\o{f},\frac{A_{41}}{B_{41}})=\frac{P_{11,2}}{(B_{41})^2}$, where $P_{11,2}(\o{f}, \o{g})$ is a polynomial of degree $(11,2)$
vanishing at the 8 points with multiplicity 2.
Since $F(\o{f}, g(\o{f},\o{g}))|_{\o{f}=\o{f}(u_i)}=F(\o{f}(u_i),g(u_i))=0$ (regardless of $\o{g}$), the polynomial $P_{11,2}$ 
is factorized as $P_{11,2}= \tilde{F}_{32}(\o{f}, \o{g}) \prod_{i=1}^8 \{\o{f}-\o{f}(u_i)\}$ and $\tilde{F}_{32}=0$ gives the 
desired curve of degree $(3,2)$ passing through the 8 points.\footnote{
In terms of the Picard lattice, this part corresponds to the fact that
$3H_1+2H_2-\sum_{i=1}^8 E_i$ is invariant under the substitution
$H_2 \mapsto 4H_1+H_2-\sum_{i=1}^8 E_i$, $E_i \mapsto H_1-E_i$ ($i=1,\ldots,8$).}
The other two vanishing conditions follow from
the relation $\tilde{F}_{32}(\o{f}(u), \o{g}(u))=0 \Leftrightarrow
F(\o{f}(u)=\o{f}(\frac{h_1}{q u}), g(\frac{h_1}{q u}))=0$ which holds for generic $u(\neq u_i)$.\qed

\begin{lemma}\label{lem:10pt-cond}
The equation $L_{1u}=0$ gives a curve of degree $(3,2)$ in $(\o{f},\o{g})$ 
passing through the 10 points $(\o{f},\o{g})=(\o{f}(u),\o{g}(u))$ with $u=u_1, \ldots, u_8, z, \frac{h_1}{z}$.
\end{lemma}
\noindent
{\it Proof.}
It follows from the Lemma \ref{lem:L1u} and Lemma \ref{lem:gog}. \qed

Lemma \ref{lem:10pt-cond} ensures the vanishing properties of $L_{1u}$
which must be satisfied by $T$-evolution of $L_1$ at the first 10 points. 
In order to prove the main theorem, it is enough to prove the following
\begin{lemma}
The equation $L_{1u}$ vanishes at $\o{Q}(z)$ and $\o{Q}(\frac{z}{q})$, where
$\o{Q}(u)=(\o{f},\o{g})$ is defined by $\o{f}=\o{f}(u)$ and $\dfrac{\o{g}-\o{g}(u)}{\o{g}-\o{g}(\frac{h_1}{qu})}=\dfrac{\o{Y}(qu)}{\o{Y}(u)}$,  
for $u=z, \frac{z}{q}$.
\end{lemma}
\noindent
{\it Proof.} 
When $\o{f}=\o{f}(\frac{z}{q})$, from eq.(\ref{eq:gueq}) and eq.(\ref{eq:Prel2}), we have
\begin{equation}
\begin{array}{rl}
\dfrac{g-g(\frac{z}{q})}{g-g(\frac{h_1}{z})}&=\dfrac{\o{g}-\o{g}(\frac{h_1}{z})}{\o{g}-\o{g}(\frac{z}{q})}
\dfrac{h_1 q}{z^2}\dfrac{P_d(\frac{h_1}{q},\o{f}(\frac{z}{q}))+(\frac{h_1}{q})^3(\frac{z}{q})^2 P_n(\frac{h_1}{q},\o{f}(\frac{z}{q}))}
{P_d(\frac{h_1}{q},\o{f}(\frac{z}{q}))+(\frac{h_1}{q})^3(\frac{h_1}{z})^2 P_n(\frac{h_1}{q},\o{f}(\frac{z}{q}))}\\[4mm]
&=\dfrac{\o{g}-\o{g}(\frac{h_1}{z})}{\o{g}-\o{g}(\frac{z}{q})}\dfrac{h_1^4 q^4}{z^8}\dfrac{U(\frac{z}{q})}{U(\frac{h_1}{z})}.
\end{array}
\end{equation}
Then, from the residue of eq.(\ref{eq:L1up}) at $\o{f}=\o{f}(\frac{z}{q})$, we have
\begin{equation}
\dfrac{\o{Y}(\frac{z}{q})}{\o{Y}(z)}=\dfrac{z^8}{h_1^4q^4}\dfrac{U(\frac{h_1}{z})}{U(\frac{z}{q})}\dfrac{g-g(\frac{z}{q})}{g-g(\frac{h_1}{z})}
=\dfrac{\o{g}-\o{g}(\frac{h_1}{z})}{\o{g}-\o{g}(\frac{z}{q})}.
\end{equation}
This is the desired relation for $u=\frac{z}{q}$. The relation for $u=z$ is similar.\qed

The proof of the main theorem is completed.

\section{Degenerations}\label{sect:dege}

In this section, we consider the degeneration limit of the $E^{(1)}_8$ system
to $E^{(1)}_7$, $E^{(1)}_6$, $D^{(1)}_5$. For the corresponding $q$-Painlev\'e equations
see \cite{KMNOY2}, \cite{Tsuda2} and references therein.

\subsection{Degeneration from $E^{(1)}_8$ to $E^{(1)}_7$}

We put $(h_1, h_2)=(t \epsilon, \frac{\epsilon}{t})$ and 
$(u_1, \ldots, u_8)=(b_1,\ldots,b_4,\frac{\epsilon}{b_5}, \ldots, \frac{\epsilon}{b_8})$,
hence $q=\frac{b_5b_6b_7b_8}{b_1b_2b_3b_4}$. Under the limit $\epsilon \rightarrow 0$ we have
\begin{equation}
\begin{array}{ll}
P_n(h_2, g) \rightarrow g^4 B_1(\frac{1}{g}), &
P_d(h_2, g) \rightarrow \epsilon^4 \frac{g^4}{q} B_2(\frac{1}{tg}), \\[4mm]
P_n(\frac{h_1}{q}, \o{f}) \rightarrow  \o{f}^4 B_1(\frac{1}{\o{f}}), &
P_d(\frac{h_1}{q}, \o{f}) \rightarrow \epsilon^4 \frac{\o{f}^4}{q} B_2(\frac{t}{q \o{f}}), \\[4mm]
U(\frac{z}{q}) \rightarrow (\frac{z}{q})^8 B_1(\frac{q}{z}), &
U(\frac{h_1}{z}) \rightarrow \epsilon^4 \frac{1}{q} B_2(\frac{t}{z}),
\end{array}
\end{equation}
where
\begin{equation}
B_1(z)=\prod_{i=1}^4(1-b_i z),\quad
B_2(z)=\prod_{i=5}^8(1-b_i z).
\end{equation}
In addition to this limiting procedure, we also make a change of coordinate : $g \rightarrow \frac{1}{g}$.
Then the configuration of the 8 points are
$(f,g)= (b_i,\frac{1}{b_i})_{i=1,\ldots,4}$, and $(b_i t,\frac{t}{b_i})_{i=5,\ldots,8}$.
They are on the curve $(fg-1)(fg-t^2)=0$.

The $q$-Painlev\'e equation of type $E^{(1)}_7$ is then given by  $(b_i,t,f,g) \mapsto (b_i,\frac{t}{q},\o{f},\o{g})$, where
\begin{equation}
\begin{array}l
\dfrac{(fg-1)(\o{f}g-1)}{(fg-t^2)(\o{f}gq-t^2)}=\dfrac{B_1(g)}{t^4 B_2(\frac{g}{t})}, \\[5mm]
\dfrac{(\o{f}g-1)(\o{f}\o{g}-1)}{(\o{f}g-t^2)(\o{f}\o{g}q^2-t^2)}=\dfrac{B_1(\frac{1}{\o{f}})}{q^3 B_2(\frac{t}{\o{f}q})}.
\end{array}
\end{equation}

The Lax pair is
\begin{equation}
\begin{array}{rl}
L_1=&\dfrac{(1-t^2)}{g z^2}\left[\dfrac{q {B_1}(g)}{(f g-1) (g z-q)}
-\dfrac{t^4 {B_2}(\frac{g}{t})}{(f g-t^2) (g z-t^2)}\right] Y(x)\\[6mm]
&+\dfrac{{B_2}(\frac{t}{z})}{t^2(f-z)}\left[Y(qz)-\dfrac{t^2(1-gz)}{t^2-gz}Y(z)\right]\\[6mm]
&+\dfrac{t^2{B_1}(\frac{q}{z})}{q(fq-z)}\left[Y(\frac{z}{q})-\dfrac{q t^2-gz}{t^2(q-gz)}Y(z)\right]=0,
\end{array}
\end{equation}
and
\begin{equation}
L_2=\frac{qt^2 - gz}{t^2}Y(z)+ (gz-q)Y(\frac{z}{q})+ \frac{gz(fq - z)}{q^2}\o{Y}(\frac{z}{q})=0.
\end{equation}

\subsection{Degeneration from $E^{(1)}_7$ to $E^{(1)}_6$}

Degeneration from $E^{(1)}_7$ to $E^{(1)}_6$ is obtained by putting 
$b_5 \rightarrow b_5/\epsilon$, $b_6 \rightarrow b_6/\epsilon$, $b_7 \rightarrow b_7 \epsilon$, $b_8 \rightarrow b_8 \epsilon$
and $t \rightarrow t \epsilon$ and taking the limit $\epsilon \rightarrow 0$.

The 8 points configuration:
$(f,g)= (b_i,\frac{1}{b_i})_{i=1,\ldots,4}$, $(b_i t,0)_{i=5,6}$
and $(0,\frac{t}{b_i})_{i=7,8}$.

The $q$-Painlev\'e equation:
\begin{equation}
\begin{array}l
\dfrac{(f g-1) (\o{f} g-1) }{f \o{f} }=\dfrac{q(b_{1} g-1) (b_{2} g-1) (b_{3} g-1) (b_{4}g-1)}{b_5 b_6(b_{7} g-t) (b_{8}g-t)},\\[6mm]
\dfrac{(\o{f} g-1) (\o{f} \o{g}-1)}{ g \o{g}}=\dfrac{q^2 (b_{1}-\o{f}) (b_{2}-\o{f}) (b_{3}-\o{f}) (b_{4}-\o{f})}
{(\o{f} q-b_{5} t) (\o{f} q-b_{6} t)}.
\end{array}
\end{equation}

The Lax pair:
\begin{equation}
\begin{array}{rl}
L_1=&\dfrac{(b_{1} q-z) (b_{2} q-z) (b_{3} q-z) (b_{4} q-z) t^2}{q (f q-z) z^4}\Big[Y(\frac{z}{q})-\dfrac{g z}{t^2(g z-q)}Y(z)\Big]\\[5mm]
&+\Big[\dfrac{(b_{1} g-1)(b_{2} g-1) (b_{3} g-1) (b_{4} g-1) q }{g (f g-1) z^2 (g z-q)}
-\dfrac{b_5b_6(b_{7} g-t) (b_{8}g-t)}{f g z^3}\Big]Y(z)\\[5mm]
&+\dfrac{(b_{5} t-z) (b_{6} t-z)}{(f-z) z^2 t^2} \Big[Y(q z)-\dfrac{(g z-1) t^2 }{g z}Y(z)\Big]=0,\\[6mm]
L_2=&\dfrac{g z}{t^2}Y(z)+(q-g z) Y(\frac{z}{q})-\dfrac{g z(f q-z)}{q^2}\o{Y}(\frac{z}{q})=0.
\end{array}
\end{equation}

A $2\times 2$ matrix Lax formalism for the $q$-Painlev\'e equation with $E^{(1)}_6$-symmetry has been obtained by Sakai \cite{SakaiA2}.
The scalar Lax equation here may be equivalent to Sakai's one, though we could not confirm it so far.
   
\subsection{Degeneration from $E^{(1)}_6$ to $D^{(1)}_5$}

Degeneration from $E^{(1)}_6$ to $D^{(1)}_5$ is obtained by putting 
$b_1 \rightarrow b_1/\epsilon$, $ b_2 \rightarrow b_2/\epsilon$, $ b_3 \rightarrow b_3 \epsilon$, $ b_4 \rightarrow b_4 \epsilon$, $
f \rightarrow f \epsilon$, $ g \rightarrow g \epsilon$, $ t \rightarrow t \epsilon$, $ z \rightarrow z \epsilon$, 
$\o{Y} \rightarrow \epsilon^{-3} \o{Y}$, and taking the limit $\epsilon \rightarrow 0$. 

The 8 points configuration:
$(f,g)= (\infty,\frac{1}{b_i})_{i=1,2}$, $(b_i,\infty)_{i=3,4}$, $(b_i t,0)_{i=5,6}$
and $(0,\frac{t}{b_i})_{i=7,8}$.

The $q$-Painlev\'e equation:
\begin{equation}
\begin{array}l
f \o{f}=\dfrac{b_5b_6(b_{7} g-t) (b_{8}g-t)}{q(b_{1} g-1) (b_{2} g-1)},\\[6mm]
g \o{g}=\dfrac{(\o{f} q-b_{5} t) (\o{f}q-b_{6} t)}{q^2 b_{1} b_{2} (b_{3}-\o{f}) (b_{4}-\o{f})}.
\end{array}
\end{equation}

The Lax pair:
\begin{equation}
\begin{array}{rl}
L_1=&\dfrac{b_{1} b_{2} q (b_{3} q-z) (b_{4} q-z)  t^2}{(f q-z)z^2}\Big[Y(\frac{z}{q})+\dfrac{q z}{q t^2}Y(z)\Big]\\[5mm]
&+\Big[\dfrac{(b_{1} g-1) (b_{2} g-1)}{g}
-\dfrac{b_{5} b_{6}(b_{7} g-t) (b_{8} g-t)}{f g z}\Big]Y(z)\\[5mm]
&+\dfrac{(b_{5} t-z) (b_{6} t-z)}{(f-z) t^2}\Big[Y(q z)+\dfrac{t^2 }{g z}Y(z)\Big]=0,\\[6mm]
L_2=&\dfrac{g z}{t^2} Y(z)+qY(\frac{z}{q})-\dfrac{gz(fq-z)}{q^2}\o{Y}(\frac{z}{q})=0.
\end{array}
\end{equation}

This result is essentially equivalent to the original $2 \times 2$ construction by Jimbo-Sakai \cite{JS}. 
Further degenerations of this has been comprehensively studied by Murata \cite{Murata2}.
The Lax formalism of Jimbo-Sakai is also derived from $q$-KP/UC hierarchy, and this method
is also effective to the higher order generalizations \cite{Tsuda}.

\appendix
\section{Weyl group actions}

Here, we will discuss the affine Weyl group symmetry of the $q$-Painlev\'e equation (\ref{eq:fueq}) (\ref{eq:gueq}),
based on the constructions in \cite{KMNOY1},\cite{Murata} and \cite{Tsuda2}.

Define multiplicative transformations $s_{ij}$, $c$, $\mu_{ij}$, $\nu_{ij}$ ($1\leq i\neq j \leq 8$)
acting on variables  $h_1, h_2, u_1, \ldots, u_8$ as
\begin{equation}\label{eq:pic-action}
\begin{array}l
s_{ij} = \{u_i \leftrightarrow u_j\},\qquad
c= \{h_1 \leftrightarrow h_2\},\\
\mu_{ij} = \{h_1 \mapsto \dfrac{h_1h_2}{u_iu_j}, \quad
u_i \mapsto \dfrac{h_2}{u_j}, \quad u_j \mapsto \dfrac{h_2}{u_i}\},\\
\nu_{ij} = \{h_2 \mapsto \dfrac{h_1h_2}{u_iu_j}, \quad
u_i \mapsto \dfrac{h_1}{u_j}, \quad u_j \mapsto \dfrac{h_1}{u_i}\}.
\end{array}
\end{equation}
These actions generate the affine Weyl group of type $E^{(1)}_8$.
A choice of simple reflections is
\begin{equation}
\begin{array}{cccccccccccccccccc}
&&&&s_{12}\\
&&&&\vert\\
c&-&\mu_{12}&-&s_{23}&-&s_{34}&-&\cdots&-&s_{78} &&.
\end{array}
\end{equation}
The transformations (\ref{eq:pic-action}) naturally act on 
the 8 points configuration $(f_i,g_i)=(u_i+\frac{h_1}{u_i}, u_i+\frac{h_2}{u_i})$ as
$\mu_{12}(f_1)=\frac{h_1}{u_1}+\frac{h_2}{u_2}$, $\mu_{12}(g_1)=u_1+\frac{h_2}{u_1}$ for example.
Then one can extend the Weyl group actions bi-rationally including generic variables $(f,g) \in \P^1\times \P^1$.
The nontrivial actions are as follows:
\begin{equation}
\begin{array}l
c(f)=g, \quad c(g)=f, \quad
\mu_{ij}(f)=\tilde{f},\quad
\nu_{ij}(g)=\tilde{g},
\end{array}
\end{equation}
where, $\tilde{f}$ and $\tilde{g}$ are rational functions in $(f,g)$ defined by
\begin{equation}
\begin{array}l
\dfrac{\tilde{f}-\mu_{ij}(f_i)}{\tilde{f}-\mu_{ij}(f_j)}=\dfrac{(f-f_i)(g-g_j)}{(f-f_j)(g-g_i)},\\[6mm]
\dfrac{\tilde{g}-\nu_{ij}(g_i)}{\tilde{g}-\nu_{ij}(g_j)}=\dfrac{(g-g_i)(f-f_j)}{(g-g_j)(f-f_i)}.
\end{array}
\end{equation}

Let $r$ and $T_1$ be the following compositions
\begin{equation}
\begin{array}l
r=s_{12}\mu_{12}s_{34}\mu_{34}s_{56}\mu_{56}s_{78}\mu_{78}, \quad
T_1=crcr.
\end{array}
\end{equation}
Their actions on variables $(h_i, u_i, f, g)$ are given by
\begin{equation}
\begin{array}{lllll}
r(h_1)=v h_2, &
r(h_2)=h_2, &
r(u_i)=\dfrac{h_2}{u_i},&
r(f)=\o{f} v, & r(g)=g, \\
T_1(h_1)=\frac{h_1}{q} v^2, &
T_1(h_2)=q h_2 v^2, &
T_1(u_i)=u_i v, &
T_1(f)=\o{f} v, &
T_1(g)=\o{g} v,
\end{array}
\end{equation}
where $v=qh_2/h_1$.
Hence, the evolution $T$ of $q$-Painlev\'e equation (\ref{eq:T-bar})
is the affine Weyl group translation $T_1$ up to re-scaling of the parameters and variables.

\vskip10mm
\noindent
{\bf Acknowledgment.} 
The author would like to thank N.Joshi, K.Kajiwara and T.Tsuda for their interests and
discussions. This work is partly supported by JSPS KAKENHI No.21340036 and S-19104002.

\end{document}